\documentclass[aps,pra,onecolumn,8pt,showpacs]{revtex4}
\usepackage{mathrsfs}
\usepackage{graphicx}

\usepackage{amsmath}
\usepackage{amssymb}
\usepackage{CJK}
\usepackage{hyperref}
\usepackage{subfigure}
\usepackage{epsfig}

\usepackage{graphicx}
\usepackage{dcolumn}
\usepackage{bm}
\usepackage{amsmath}
\usepackage{amssymb}
\usepackage{latexsym}
\usepackage{epsfig}
\usepackage{amsbsy}
\usepackage{array}
\usepackage{amssymb}
\usepackage{setspace}
\usepackage{bm}
\usepackage{epstopdf}
\usepackage{subfigure}
\usepackage{color}
\usepackage{xcolor}

\begin{document}

\title{Strong coupling between two spin ensembles via a large Josephson junction}

\author{Feng-Yang Zhang$^{1}$\footnote{E-mail:zhangfy@dlnu.edu.cn}, Wen-Lin Li$^2$, and Chui-Ping Yang$^3$\footnote{E-mail:yangcp@hznu.edu.cn}}
 \affiliation{$^1$School of Physics and Materials Engineering, Dalian Nationalities University, Dalian $116600$, China \\$^2$School of Physics
and Optoelectronic Technology, Dalian University of Technology, Dalian 116024, China\\
$^3$Department of Physics, Hangzhou Normal University, Hangzhou, Zhejiang 310036, China}

\begin{abstract}
We propose a method to achieve strong coupling between a spin ensemble and a large Josephson junctions (LJJ). Then, the strong coupling between two spin ensembles can be induced by a LJJ. A non-adiabatic holonomic single-qubit quantum gates is realized. Moreover, with the dispersive interaction between the spin ensembles and the LJJ, the high-fidelity two-qubit phase gate can be implemented with two spin ensembles within an operation time $0.41$ ns and $\pi$ phase-swap gate can be realized within an operation time $103.11$ ns with a high fidelity greater than $99\%$.
\end{abstract}
  \maketitle
\newpage
\section{Introduction}
Due to the sufficiently long electronic spin lifetime as
well as the possibility of coherent manipulation at room
temperature \cite{e}, the nitrogen-vacancy (NV) center in diamond
provides an arena to study various macroscopic quantum phenomena and
acts as a perfect candidate toward quantum computers. Experimentally,
with NV centers Deutsch-Jozsa quantum algorithm  \cite{f}, quantum memory \cite{f1}, quantum logical NOT and a conditional two-qubit gate \cite{f2}, and decoherence-protected quantum gates for the electron-nuclear spin register \cite{f3}, electron spin resonance detected by a superconducting qubit \cite{f33}, and controling spin relaxation \cite{f34} have been realized, respectively. Theoretically, a multiqubit conditional phase gate with three NV centers coupled to a whispering-gallery
mode cavity has been proposed \cite{f4}, quantum-information transfer with NV centers coupled to a whispering-gallery
microresonator has been proposed \cite{g}, and anomalous decoherence
effect has been found in NV center \cite{h}.

Recently, the hybrid quantum system consisting of a flux qubit and a NV ensemble has been proposed \cite{ii}. The coupling strength between the flux qubit and the NV ensemble is stronger than that between NV centers and a
transmission line resonator \cite{ii}. And, the strong coupling \cite{ii1}, quantum information transfer \cite{ii2}, as well as observation of dark states \cite{ii3} have been demonstrated in this hybrid system. Also, many theoretical works have been done in the quantum information based on this hybrid system. For example, Refs. \cite{ii4, ii5} proposed the quantum information transfer between two spatially-separated NV ensembles. H\"{u}mmer \emph{et al.} \cite{ii55} proposed how to simulate a localization-delocalization transition based on an array of superconducting flux qubits which are coupled to a diamond crystal containing
NV centers. Qiu \emph{et al.} \cite{ii6} showed that the coupling strength
between flux qubits and NV ensembles can reach the strong and even ultrastrong coupling regimes
by either engineering the hybrid structure in advance or tuning the excitation frequencies of NV
ensembles via external magnetic fields. L\"{u} \emph{et al.} \cite{ii7} proposed how to realize high-fidelity quantum storage using two coupled
flux qubits and a NV ensemble. Song \emph{et al.} \cite{ii8} studied a scheme for creating macroscopic entangled coherent states of separate NV ensembles that are coupled to a flux qubit.

\begin{figure}
\includegraphics[scale=0.5]{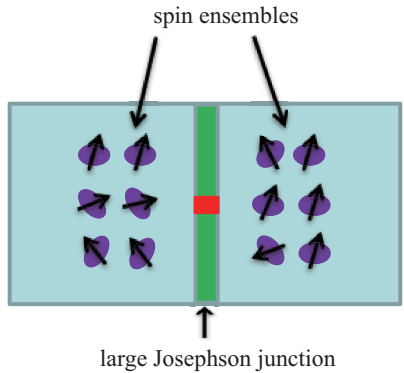}
\caption{(Color online) Schematic diagram of the proposed hybrid quantum system, which consists of a large Josephson junction and two spin ensembles.} 
\label{fi}
\end{figure}

In this paper, we study a hybrid quantum system, which
consists of the NV ensembles and a large Josephson junctions (LJJ), as shown in Fig.\ref{fi}. The key point of our scheme is that the strong coupling between the LJJ and the NV ensembles can be achieved. And  a non-adiabatic holonomic single-qubit quantum gates is implemented. Also, in the large detuning regime between LJJ and NV ensembles, the LJJ can induce the strong interaction of two spatially-separated NV ensembles. Considering decoherence
in experimentally available systems, we show the feasibility of
achieving high-fidelity quantum logic gates.

\section{system and model}
We start from the simple structure consisting of a NV ensemble and a LJJ. The NV ensemble is realized by NV centers with number \emph{N}.
Each NV center consists of a nitrogen impurity in the diamond lattice
with a vacancy on a  neighbouring lattice site. The ground state of a NV center has a spin one, with the sublevels $m_{s}=0$ and $m_{s}=\pm1$ separated by zero-field splitting $D_{gs}$. The NV center can be described by the
Hamiltonian \cite{l1, l2}
\begin{eqnarray}
H_{NV}=D_{gs}S_{z}^{2}+E(S_{x}^{2}-S_{y}^{2})+g_{e}\mu_{B}\textbf{B}\bullet\textbf{S},
\end{eqnarray}
where $E$ is the strain-induced splitting coefficient, $S_{x},~ S_{y}$, and $S_{z}$ are the components
of $\textbf{S}$ which denote the Pauli spin-one operators, $g_{e}=2$ is the NV Land\'{e}
factor, $\mu_{B}=14$MHz mT$^{-1}$ is the Bohr magneton, and $\textbf{B}$ is the applied magnetic field.
In this paper, the quantum information is encoded in sublevels $|m_{s}=0\rangle\equiv|a\rangle$ and $|m_{s}=\pm1\rangle\equiv|b\rangle$ serving as two logic states of a qubit. For a NV ensemble is composed of NV centers with number $N$, the ground state of a NV ensemble is defined as $|0\rangle=|a_{1}a_{2}\ldots a_{N}\rangle$ while the excited state is defined
as $|1\rangle=\sigma^{+}|0\rangle=(1/\sqrt{N})\sum_{k=1}^{N}|a_{1}\ldots b_{k}\ldots
a_{N}\rangle$ (all spins are in the ground state except the \emph{k}-th spin) with operator $\sigma^{+}=(\sigma^{-})^{\dag}=(1/\sqrt{N})\sum_{k=1}^{N}|b\rangle_{k}\langle a|$. The operator $\sigma^{+}$ can create
symmetric Dicke excitation states. Thus, the Hamiltonian describing a NV ensemble reads \cite{k}
\begin{equation}
H_{NVE}=\frac{\omega_{10}}{2}\sigma_{z},
\end{equation}
where $\omega_{10}$ is the energy difference between
the lowest two levels with $|0\rangle$ and $|1\rangle$, and the operator $\sigma_{z}=|1\rangle\langle1|-|0\rangle\langle0|$ expresses the collective
spin operator for the NV ensemble.

The quantum characters of LJJ have been widely studied in the early years \cite{kk1, kk2}. The idea of using a LJJ coupled to two charge qubits was first proposed in Ref. \cite{kk3}. Then, generation of entanglement of two charge-phase qubits through a LJJ was discussed \cite{kk4}. Also, generation of macroscopic entangled coherent states with LJJs was proposed \cite{kk5}. The Hamiltonian of the LJJ can
be written as \cite{kk6} $H_{J}=E_{C}N^{2}-E_{J}\cos\gamma$,
where $E_{C}$ expresses the charging energy, $N$ is the excess Cooper pairs, $E_{J}$ denotes the Josephson energy, and $\gamma$ defines the phase drop across the LJJ.
When the LJJ works in the phase regime, one can use a harmonic
oscillator model to characterize the LJJ. Thus, the Hamiltonian for the LJJ is \cite{kk6}
\begin{eqnarray}
H_{J}=\omega a^{\dag}a,
\end{eqnarray}
with bosonic operators $a^{\dag}=\frac{\xi}{2}\gamma-i\frac{1}{2\xi}N$ and $a=\frac{\xi}{2}\gamma+i\frac{1}{2\xi}N$;
and plasma frequency $\omega=\sqrt{8E_{C}E_{J}}$. Here, $\xi=(E_{J}/E_{C})^{1/4}$.

The magnetic coupling strength between the \emph{k}-th NV center and the LJJ can be roughly estimated as $g_{k}=2g_{e}\mu_{B}B$. Here, $B$ denotes magnetic field which can be estimated using the Biot-Savart law $B\simeq\mu_{0}I_{0}/4\pi r$, where $\mu_{0}=4\pi\times10^{-7}$ Nm$^{2}$, $r$ expresses the distance between the LJJ and a NV center, and $I_{0}$ is critical current. The interaction Hamiltonian for a LJJ coupled to a NV ensemble can be represented by $\sum_{k}g_{k}(a^{\dag}|a\rangle_{k,k}\langle b|+a|b\rangle_{k,k}\langle a|)$. In our architecture (Fig.1), NV ensembles are on the left and right of the LJJ, respectively. Here, we neglect the direct interaction of the two NV ensembles. Thus, the interaction Hamiltonian of two NV ensembles coupled to a LJJ can be written as
\begin{eqnarray}
H_{int}=\sum_{j=1}^{2}G_{j}a\sigma^{+}_{j}+\texttt{H.c.}
\end{eqnarray}
where $G_{j}$ is the collective coupling constant. Below, we give an estimation on this coupling strength $G_{j}=\sqrt{N}g_{k}$. According to experiments \cite{kk2}, the critical current $I_{0}$ can be chosen as $\sim21\mu$A. The distance $r$ from the center of the ensemble to a superconducting qubit is $\sim1.2\mu$m \cite{ii1}. Given these parameters, we can
estimate the coupling strength as $g_{k}\sim0.62$MHz. The number $N$ of NV centers takes $\sim10^{6}$, the collective coupling constant is $G_{j}\sim620$MHz. The decoherence rate of the NV ensemble $\Gamma\sim1$MHz \cite{kk6} and the decay rate of the LJJ $\kappa\sim3.3$MHz \cite{kk2} have been reported. Thus, this coupling is in the strong coupling regime.

\section{generation of entanglement}
\begin{figure}
  \includegraphics[scale=0.35]{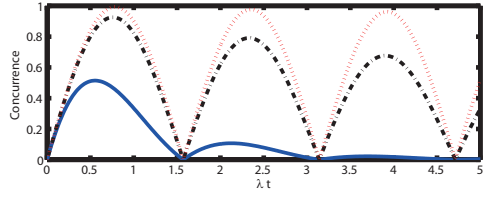}\\
  \caption{(Color online) The time evolution of the concurrence of the state $|\Psi(t)\rangle$ for different spontaneous rates. Solid-blue lines, dot-black lines, and dash-red lines correspond to $\Gamma=\lambda$, $\Gamma=0.1\lambda$, and $\Gamma=0.01\lambda$, respectively.} \label{f1}
\end{figure}

We assume the system work within the large detuning condition $\delta\gg
G_{i}$, where $\delta=\omega-\omega_{10}$ is the detuning between the LJJ and the NV ensemble. There is no energy exchange between the LJJ and each NV ensemble. The indirect interaction of the two NV ensembles can be induced by a LJJ without excitation.
If we assume the LJJ is initially in the vacuum state, the
effective Hamiltonian is given by
\begin{eqnarray}
H_{e}=\sum_{j=1,2}\frac{G^{2}_{j}}{\delta}|1\rangle_{j,j}\langle 1|+\frac{G_{1}G_{2}}{\delta}(\sigma^{+}_{1}\sigma^{-}_{2}+\sigma^{-}_{1}\sigma^{+}_{2}). \label{1}
\end{eqnarray}
The first term describes the energy levels shifts, the
last term describes the coupling of the two separated NV ensembles. In Ref. \cite{m}, a Hamiltonian similar to Eq. (\ref{1}) was proposed using a cavity coupled to two atoms. We now consider the spontaneous emission of the NV ensembles. Under the assumption of weak decay of the LJJ, the evolution of the system is governed by
the conditional Hamiltonian
\begin{eqnarray}
H_{c}=H_{e}-i\frac{\Gamma}{2}\sum_{j=1,2}|1\rangle_{j,j}\langle 1|.
\end{eqnarray}
 Suppose
the system is initially prepared in the state $|0\rangle_{1}|1\rangle_{2}$ and the coupling strength $G_{j}$ is equal, i.e. $G_{1}=G_{2}=G$, then
the state evolution of the system is given by
\begin{eqnarray}
|\Psi(t)\rangle=C_{1}(t)|0\rangle_{1}|1\rangle_{2}+C_{2}(t)|1\rangle_{1}|0\rangle_{2}, \label{2}
\end{eqnarray}
with the coefficients $C_{1}(t)=\frac{1}{2}\exp(-\Gamma t/2)[1+\exp(-i2\lambda t)]$ and $C_{2}(t)=\frac{1}{2}\exp(-\Gamma t/2)[\exp(-i2\lambda t)-1]$ with the parameter $\lambda=G^{2}/\delta$, which denotes the effective coupling strength of two NV ensembles. The parameter $\lambda$ can be estimated as $34.6$MHz, which shows that in the large-detuning regime, the strong coupling between two NV ensembles, mediated by a LJJ, can be obtained. Eq. (\ref{2}) expresses the entangled state of two NV ensembles. For different spontaneous rates $\Gamma$, we plot the concurrence change with $\lambda t$ in Fig.\ref{f1}. Obviously, concurrence can reach the maximum at some moment with a low spontaneous rate.
In other words, the LJJ can induce maximal entanglement of two NV ensembles.
\section{realizing single-qubit logic gates}
Recently, holonomic quantum computation based on nonadiabatic non-Abelian geometric phases has been proposed \cite{w1}. These kinds of geometric gates have been investigated in different physical systems \cite{w2, w3, w4, w5, w6, w7}. We can construct the universal set of
non-adiabatic holonomic single-qubit quantum gates based on Hamiltonian (4).
We assume that the system is initially in the state $|100\rangle\equiv|1\rangle_{1}\otimes|0\rangle_{L}\otimes|0\rangle_{2}$, i.e., they denote the states of the left NV ensemble, the LJJ and the right NV ensemble, respectively. Under the Hamiltonian (4), the state of the system evolves in
the subspace $\{|\varphi_{1}\rangle=|100\rangle, |\varphi_{2}\rangle=|010\rangle, |\varphi_{3}\rangle=|001\rangle\}$. In such a subspace the eigenstates of the system are
\begin{eqnarray}
|\psi_{d}\rangle=\frac{1}{\mathcal{G}}(-G_{1}|\varphi_{1}\rangle+G_{2}|\varphi_{3}\rangle),
\end{eqnarray}
\begin{eqnarray}
|\psi_{\pm}\rangle=\frac{1}{\sqrt{2}}\left[\frac{1}{\mathcal{G}}(G_{2}|\varphi_{1}\rangle+G_{1}|\varphi_{3}\rangle)\pm|\varphi_{2}\rangle\right],
\end{eqnarray}
with the corresponding eigenenergies given by $E_{d}=0$ and $E_{\pm}=\pm\sqrt{G^{2}_{1}+G^{2}_{2}}=\mathcal{G}$. $|\psi_{d}\rangle$ expresses the dark state, which is decoupled from the Hamiltonian and undergoes no transition during the application of the
driving fields. The bright state is $|\psi_{b}\rangle=\frac{1}{\mathcal{G}}(G^{*}_{2}|\varphi_{1}\rangle+G^{*}_{1}|\varphi_{3}\rangle)$. It is obvious that the effective Rabi frequency between the bright state $|\psi_{b}\rangle$ and the quantum state $|\varphi_{2}\rangle$ is $\mathcal{G}$. Without loss of generality, we set $G_{1}/\mathcal{G}=\cos(\theta/2)$ and $G_{2}/\mathcal{G}=e^{i\phi}\sin(\theta/2)$, with $0\leq\theta\leq\pi$ decided by
the ratio between the coupling strength $G_{1}$ and $G_{2}$, and $0\leq\phi\leq2\pi$ depending on the relative phase. When condition $\int_{0}^{T}\mathcal{G}dt=\pi$
is satisfied, the dark state and bright state
undergo a cyclic evolution.
In the computational basis $\{|\varphi_{1}\rangle, |\varphi_{3}\rangle\}$, the
final evolution operator is
\begin{equation}
\left(
\begin{array}{cc}
  -\cos\theta & \sin\theta e^{i\phi} \\
  \sin\theta e^{-i\phi} & \cos\theta
\end{array}
\right),
\end{equation}
which can be used to realize any single-qubit rotation.
\section{realizing two-qubit logic gates}
Two-qubit logic gates play a key role in quantum computation and quantum information. A multiqubit gate can be composed of single-qubit and two-qubit gates. In this section, we discuss how to realize a two-qubit phase gate and a two-qubit $\pi$ phase-swap gate with two spin ensembles, respectively.

\subsection{phase gate}
In order to implement a phase gate, an external signal is applied to
the LJJ via an on-chip antenna. The Hamiltonian of the external signal driving the LJJ can be modeled by \cite{n}
\begin{equation}
H_{d}=\varepsilon(a^{\dag}e^{-i\omega_{d}t}+ae^{i\omega_{d}t}),
\end{equation}
where parameters $\varepsilon$ and $\omega_{d}$ express the amplitude and
the frequency of the external signal, respectively. Then, in the Schr\"{o}dinger picture, the total Hamiltonian of
the system can be written as
\begin{eqnarray}
H_{t}&=&H_{J}+H_{NVE}+H_{int}+H_{d} \nonumber\\
&=&\omega a^{\dag}a+\frac{\omega_{10}}{2}\sigma_{z}+\sum_{j=1}^{2}G_{j}(a\sigma^{+}_{j}+a^{\dag}\sigma^{-}_{j}) \nonumber\\
&&+\varepsilon(a^{\dag}e^{-i\omega_{d}t}+ae^{i\omega_{d}t}).
\end{eqnarray}

We introduce a displacement-transformation operator $D(\alpha)=\exp(\alpha
a^{\dag}-\alpha^{*}a)$, where $\alpha$ is a complex number.
After the displacement-transformation $D(\alpha)$ of the Hamiltonian (9), we obtain a new Hamiltonian
$H_{T}=D^{\dag}(\alpha)H_{t}D(\alpha)-iD^{\dag}(\alpha)\dot{D}(\alpha)$, where the expression of complex number is $\dot{\alpha}=-i\omega\alpha-i\varepsilon e^{-i\omega_{d}t}$. The drive amplitude $\varepsilon$ is independent of
time. A rotating frame transformation
$U_{R}=\exp[-i\omega_{d}(\sigma_{z}^{j}+a^{\dag}a)t]$ is applied to Hamiltonian $H_{T}$, then in the interaction picture, we take $\omega_{10}=\omega_{d}$, then the $H_{T}$
becomes \cite{d1}
\begin{eqnarray}
H_{T}=\sum_{j=1,2}\left[\Omega_{j}\sigma_{x}^{j}+G_{j}(\sigma_{+}^{j}ae^{-i\delta
t}+\sigma_{-}^{j}a^{\dag}e^{i\delta t})\right],
\end{eqnarray}
where $\Omega_{j}=\varepsilon G_{j}/\delta$. The Hamiltonian (10) can be
divided two parts, which include the free term $H'_{0}=\sum_{j=1,2}\Omega_{j}\sigma_{x}^{j}$
and the interaction term $H'_{I}=\sum_{j=1,2}G_{j}(\sigma_{+}^{j}ae^{-i\delta
t}+\sigma_{-}^{j}a^{\dag}e^{i\delta t})$. In the interaction picture, we
define the new orthogonal bases
$|\pm\rangle_{j}=(|0\rangle_{j}\pm|1\rangle_{j})/\sqrt{2}$. Under
the strong driving regime i.e. $\Omega_{j}\gg \{\delta,~G_{j}\}$, we
eliminate the fast-oscillating terms and obtain effective
Hamiltonian \cite{p}

\begin{equation}
H_{eff}=\sum_{j=1,2}\frac{G_{j}}{2}\sigma^{j}_{x}(a^{\dag}e^{-i\delta
t}+ae^{i\delta t}). \label{11}
\end{equation}

If the evolution time satisfies $t=\tau_{n}=2n\pi/\delta$ for
integer $n$, the direct interaction between the
NV ensembles and a LJJ can be eliminated. Here, if we neglect a trivial universal
phase factor, the evolution operator of Eq. (\ref{11}) can be expressed as
$U(n)=\exp[-iB(n)\sigma_{x}^{1}\sigma_{x}^{2}]$ with the parameter $B(n)=-n\pi G_{1}G_{2}/\delta^{2}$. Returning to the Schr\"{o}dinger picture,
the time evolution operator is given by
\begin{eqnarray}
U(\tau)&=&e^{-iH'_{0}\tau}e^{-iB(n)\sigma_{x}^{1}\sigma_{x}^{2}},\nonumber\\
&=&e^{-i\Omega_{1}\tau \sigma_{x}^{1}}e^{-i\Omega_{2}\tau \sigma_{x}^{2}}e^{-iB(n)\sigma_{x}^{1}\sigma_{x}^{2}}.
\end{eqnarray}
If the conditions $\Omega_{1}\tau=\Omega_{2}\tau=-B(n)=\theta$ are satisfied by controlling frequencies $\Omega_{1}$ and $\Omega_{2}$,
the total time evolution operator can be rewritten as
\begin{eqnarray}
U(\theta)=\exp[-i\theta(\sigma_{x}^{1}+\sigma_{x}^{2}-\sigma_{x}^{1}\sigma_{x}^{2})].
\end{eqnarray}
Here, we choose the following bases
$\{|+\rangle_{1}|+\rangle_{2}, |+\rangle_{1}|-\rangle_{2},
|-\rangle_{1}|+\rangle_{2}, |-\rangle_{1}|-\rangle_{2}\}$, and set $4\theta=(2m+1)\pi$ (where $m$ is an integer). A two-qubit
phase gate is realized \cite{kk4, o1}
\begin{equation}
U_{CP}=\left(
        \begin{array}{cccc}
          1 & 0 & 0 & 0 \\
          0 & 1 & 0 & 0 \\
          0 & 0 & 1 & 0 \\
          0 & 0 & 0 & -1 \\
        \end{array}
      \right).
\end{equation}
According to the above parameters experimental value, we estimate the time of
implementing two-qubit phase gate. The plasma frequency the LJJ
$\omega=2\pi\times6.9$GHz \cite{kk2} and the frequency of the NV ensemble
$\omega_{10}=2\pi\times2.87$GHz \cite{k} have been reported, respectively. When we take $n=1$, the time of achieving phase gate is $\tau_{cp}\approx0.41$ns.
\subsection{$\pi$ phase-swap gate}
Here, we
choose appropriate conditions as follows: (i) two NV ensembles equally couple
to the LJJ, i.e. $G_{1}=G_{2}=G$; (ii) evolution time takes $t=\tau_{k}=\pi\delta/2G^{2}$. In the basis of two NV ensembles $\{|0\rangle_{1}|0\rangle_{2}, |0\rangle_{1}|1\rangle_{2}, |1\rangle_{1}|0\rangle_{2}, |1\rangle_{1}|1\rangle_{2}\}$, the matrix of the evolution operator of the Hamiltonian (5) reads
\begin{equation}
U_{s}= \left(
  \begin{array}{cccc}
    1 & 0 & 0 & 0 \\
    0 & 0 & 1 & 0 \\
    0 & 1 & 0 & 0 \\
    0 & 0 & 0 & -1 \\
  \end{array}
\right), \label{4}
\end{equation}
where neglect a global phase factor. This quantum logic gate can be equivalent to a $\pi$-phase gate and a swap gate. If it is realized experimentally in the future, this logic gate could be useful for quantum and
information and quantum computation.
Next, we estimate the shortest time $\tau_{k}\sim103.11$ ns for realizing this $\pi$ phase-swap gate.

\section{discussion and conclusion}
\begin{figure}
  \includegraphics[scale=0.35]{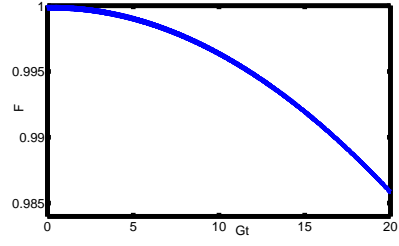}\\
  \caption{(Color online) The fidelity of the phase gate $U_{CP}$ versus the evolution time $Gt$.} \label{fcp}
\end{figure}

\begin{figure}
  \includegraphics[scale=0.4]{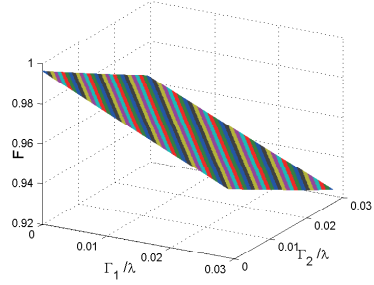}\\
  \caption{(Color online) The fidelity of the $\pi$ phase-swap gate $U_{s}$ versus the decay rates $\Gamma_{1}$ and $\Gamma_{2}$.} \label{f3}
\end{figure}

The recent experiments have reported: the electron spin
relaxation time $T_{1}$ of the NV center is $28$ s at low
temperature \cite{m1}, the dephasing time $T_{2}$ of isotopically
pure diamond sample is $2$ ms \cite{m2}, the NV ensemble coherence time $T$ approach $1$ s \cite{m22}. The time $\tau_{cp}$ and $\tau_{k}$ are much shorter than the $T_{1},~ T_{2}$ and $T$. It is necessary to
investigate the influence of decoherence of the system on the quantum logic gates. If two NV ensembles have identical relaxation rate $\Gamma_{1}$ and dephasing rate $\Gamma_{2}$, the dynamics of the lossy system is
determined by the following master equation

\begin{eqnarray}
\dot{\rho}&=&-i[H_{eff},\rho]+\frac{\kappa}{2}(2a\rho
a^{\dag}-a^{\dag}a\rho-\rho a^{\dag}a) \nonumber\\
&&+\frac{\Gamma_{1}}{2}\sum_{j=1,2}(\sigma^{j}_{z}\rho\sigma^{j}_{z}-\rho) \nonumber\\
&&+\frac{\Gamma_{2}}{2}\sum_{j=1,2}(2\sigma_{j}^{-}\rho\sigma_{j}^{+}-\rho\sigma_{j}^{+}\sigma_{j}^{-}-\sigma_{j}^{+}\sigma_{j}^{-}\rho),
\end{eqnarray}
for realizing the phase gate. Here, $\kappa$ is the decay rate of the LJJ. For achieving the $\pi$ phase-swap gate, the master equation is given by
\begin{eqnarray}
\frac{d\rho}{dt}&=&-i[H_{e},\rho]+\frac{\Gamma_{1}}{2}\sum_{j=1,2}(\sigma^{j}_{z}\rho\sigma^{j}_{z}-\rho) \nonumber\\
&&+\frac{\Gamma_{2}}{2}\sum_{j=1,2}(2\sigma_{j}^{-}\rho\sigma_{j}^{+}-\rho\sigma_{j}^{+}\sigma_{j}^{-}-\sigma_{j}^{+}\sigma_{j}^{-}\rho).
\end{eqnarray}
The fidelity is defined as \cite{w} $F=\overline{\langle\Psi|U^{\dag}\rho_{t}U|\Psi\rangle}$, where the overline indicates average overall possible initial states $|\Psi\rangle$,
$U$ is the ideal two-qubit operation, and $\rho_{t}$ is the final density operator after the $U$ operation performed in a real situation. In Fig. \ref{fcp}, we plot fidelity of the phase gate with $Gt$ for the decay rate $\kappa=\Gamma_{1}=\Gamma_{2}=1$MHz. We plot the fidelity $F$ of the $\pi$ phase-swap gate with the dephasing rate $\Gamma_{1}$ and relaxation rate $\Gamma_{2}$ in Fig.\ref{f3}. Obviously, the high-fidelity quantum logic gates can be realized when we use the previously reported decay rate $\Gamma\sim1$MHz.

Comparing with the coupling between the NV ensembles and the superconducting flux qubits \cite{ii, ii1, ii8}, our proposal is simpler in experiment. First, the preparation technology of the LJJ is much easier than that of the flux qubits. Second, the quantum information is encoded in sublevels $|m_{s}=0\rangle\equiv|a\rangle$ and $|m_{s}=\pm1\rangle\equiv|b\rangle$ serving as two logic states of a qubit, thus, there is no need of an external
magnetic field to remove the degeneracy of spin sublevels $|m_{s}=\pm1\rangle$.

In previous works on NV ensembles, the fidelity of the phase gate is about $98.23\%$ within an operation time $\sim93.87$ ns \cite{ww}.
However, in our proposal, the gate operation time is $\sim 0.41$ns and the gate can be implemented with a high-fidelity greater than $99\%$.

In summary, we have proposed a strong-coupling hybrid quantum system. With this system, the entangled state, non-adiabatic holonomic single-qubit gates, the two-qubit $\pi$ phase-swap gate and phase gate can be implemented. With the realistic
experimental parameters, we showed that the gate operation time is much shorter than the decoherence time of the system.
Hence, our scheme is implementable with the current
experimental technology.
\begin{acknowledgments}
FYZ was supported by the National Science Foundation of China under Grant Nos.
[11505024, 11447135 and 11505023], and the Fundamental Research Funds for the Central Universities
No. DC201502080407. CPY was supported in part
by the National Natural Science Foundation of China under Grant Nos. 11074062 and 11374083, and the Zhejiang Natural Science Foundation under Grant No. LZ13A040002. This work was also supported by the funds from Hangzhou City for the Hangzhou-City Quantum information and Quantum Optics Innovation Research Team. And
we acknowledge useful discussions with Zhang-qi Yin.
\end{acknowledgments}


\begin{references}

\bibitem{e} L. Childress, M. V. Gurudev Dutt, J. M. Taylor, A. S. Zibrov, F. Jelezko, J. Wrachtrup, P. R. Hemmer, and M. D. Lukin, Coherent Dynamics of Coupled Electron and Nuclear Spin Qubits in Diamond, Science \textbf{314}, 281 (2006).
\bibitem{f} F. Shi, X. Rong, N. Xu, Y. Wang, Jie Wu, B. Chong, X. Peng, J. Kniepert,
R. S. Schoenfeld, W. Harneit, M. Feng, and J. Du, Room-Temperature Implementation of the Deutsch-Jozsa Algorithm with a Single Electronic Spin in Diamond, Phys. Rev. Lett. \textbf{105}, 040504
(2010).
\bibitem{f1} G. D. Fuchs, G. Burkard, P. V. Klimov, and D. D. Awschalom, A quantum memory intrinsic to single nitrogen�Cvacancy centres in diamond, Nat. Phys. \textbf{7}, 789 (2011).
\bibitem{f2} F. Jelezko, T. Gaebel, I. Popa, M. Domhan, A. Gruber, and J.Wrachtrup, Observation of Coherent Oscillation of a Single Nuclear Spin and Realization of a Two-Qubit Conditional Quantum Gate, Phys. Rev. Lett. \textbf{93}, 130501 (2004).
\bibitem{f3} T. van der Sar, Z. H. Wang, M. S. Blok, H. Bernien, T. H. Taminiau, D. M. Toyli, D. A. Lidar, D. D. Awschalom, R. Hanson,
and V. V. Dobrovitski, Decoherence-protected quantum gates for a hybrid solid-state spin register, Nature \textbf{484}, 82 (2012).
\bibitem{f33} Y. Kubo, I. Diniz, C. Grezes, T. Umeda, J. Isoya, H. Sumiya, T. Yamamoto, H. Abe, S. Onoda, T.
Ohshima, V. Jacques, A. Dr��au, J.-F. Roch, A. Auffeves, D. Vion, D. Esteve, and P. Bertet, Electron spin resonance detected by a superconducting qubit, Phys. Rev. B \textbf{86}, 064514 (2012).
\bibitem{f34} A. Bienfait, J. J. Pla, Y. Kubo, X. Zhou, M. Stern, C. C. Lo, C. D. Weis, T. Schenkel, D. Vion, D. Esteve,
J. J. L. Morton, and P. Bertet, Controlling spin relaxation with a cavity, Nature \textbf{531}, 74 (2016).
\bibitem{f4} W. L. Yang, Z. Q. Yin, Z. Y. Xu, M. Feng, and J. F. Du, One-step implementation of multiqubit conditional phase gating with
nitrogen-vacancy centers coupled to a high-Q silica microsphere cavity, Appl. Phys. Lett. \textbf{96}, 241113 (2010).
\bibitem{g} P. B. Li, S. Y. Gao, and F. L. Li, Quantum-information transfer with nitrogen-vacancy centers coupled to a whispering-gallery microresonator, Phys. Rev. A \textbf{83}, 054306 (2011).
\bibitem{h} N. Zhao, Z. Y. Wang, and R. B. Liu, Anomalous Decoherence Effect in a Quantum Bath, Phys. Rev. Lett. \textbf{106}, 217205 (2011).
\bibitem{ii} D. Marcos, M. Wubs, J. M. Taylor, R. Aguado, M. D. Lukin, and A. S. S{\o}rensen, Coupling Nitrogen-Vacancy Centers in Diamond to Superconducting Flux Qubits, Phys. Rev. Lett. \textbf{105}, 210501 (2010).
\bibitem{ii1} X. Zhu, S. Saito, A. Kemp, K. Kakuyanagi, S. Karimoto, H. Nakano, W.J. Munro, Y. Tokura, M.S. Everitt, K. Nemoto, M. Kasu,
N. Mizuochi, K. Semba, Coherent coupling of a superconducting flux qubit to an electron spin ensemble in diamond, Nature \textbf{478}, 221 (2011).
\bibitem{ii2} S. Saito, X. Zhu, R. Ams\"{u}ss, Y. Matsuzaki, K. Kakuyanagi, T. Shimo-Oka, N. Mizuochi, K. Nemoto, W. J. Munro, K. Semba, Towards Realizing a Quantum Memory for a Superconducting Qubit: Storage and Retrieval of Quantum States, Phys.
Rev. Lett. \textbf{111}, 107008 (2013).
\bibitem{ii3} X. Zhu, Y. Matsuzaki, R. Ams\"{u}ss, K. Kakuyanagi, T. Shimo-Oka,
N. Mizuochi, K. Nemoto, K. Semba, W. J. Munro, and S. Saito, Observation of dark states in a superconductor
diamond quantum hybrid system, Nature Commu. \textbf{5}, 3524 (2014).
\bibitem{ii4} Q. Chen, W. L. Yang, and M. Feng, Controllable quantum state transfer and entanglement generation between distant nitrogen-vacancy-center ensembles coupled to superconducting flux qubits, Phys. Rev. A \textbf{86}, 022327 (2012).
\bibitem{ii5} F. Y. Zhang, C. P. Yang, and H. S. Song, Scalable quantum information transfer between nitrogen-vacancy-center ensembles, Anna. Phys. (N.Y.) \textbf{355}, 170 (2015).
\bibitem{ii55} T. H\"{u}mmer, G. M. Reuther, P. H\"{a}nggi, and D. Zueco, Nonequilibrium phases in hybrid arrays with flux qubits and nitrogen-vacancy centers, Phys. Rev. A \textbf{85}, 052320 (2012).
\bibitem{ii6} Y. Qiu, W. Xiong, L. Tian, and J. Q. You, Coupling spin ensembles via superconducting flux qubits, Phys. Rev. A \textbf{89}, 042321 (2014).
\bibitem{ii7} X. Y. L\"{u}, Z. L. Xiang, W. Cui, J. Q. You, and F. Nori, Quantum memory using a hybrid circuit with flux qubits and nitrogen-vacancy centers, Phys. Rev A \textbf{88}, 012329 (2013).
\bibitem{ii8} W. L. Song, Z. Q. Yin, W. L. Yang, X. Zhu, F. Zhou, and M. Feng, One-step generation of multipartite entanglement among nitrogen-vacancy center ensembles, Sci. Rep. \textbf{5}, 7755 (2015).
\bibitem{l1} J. H. N. Loubser, J. A. van Wyk, Electron spin resonance in the study of diamond, Rep. Progr. Phys. \textbf{41}, 1201 (1978).
\bibitem{l2} P. Neumann, R. Kolesov, V. Jacques, J. Beck, J. Tisler, A. Batalov,
L. Rogers, N. B. Manson, G. Balasubramanian, F. Jelezko, and
J. Wrachtrup, Excited-state spectroscopy of single NV defects
in diamond using optically detected
magnetic resonance, New J. Phys. \textbf{11}, 013017 (2009).
\bibitem{k} W. L. Yang, Z. Q. Yin, Y. Hu, M. Feng, and J. F. Du, High-fidelity quantum memory using nitrogen-vacancy center ensemble for hybrid quantum computation, Phys. Rev. A \textbf{84}, 010301
(2011).
\bibitem{kk1} S. Han, Y. Yu, X. Chu, S. I Chu, and Z. Wang, Time-Resolved Measurement of Dissipation-Induced Decoherence in a Josephson Junction, Science \textbf{293}, 1457 (2001).
\bibitem{kk2} J. M. Martinis, S. Nam, J. Aumentado, and C. Urbina, Rabi Oscillations in a Large Josephson-Junction Qubit, Phys. Rev. Lett. \textbf{89}, 117901 (2002).
\bibitem{kk3} J. Q. You, J. S. Tsai, F. Nori, Controllable manipulation and entanglement of macroscopic quantum states in coupled charge qubits, Phys. Rev. B \textbf{68}, 024510 (2003).
\bibitem{kk4} Y. D. Wang, P. Zhang, D. L. Zhou, and C. P. Sun, Fast entanglement of two charge-phase qubits through nonadiabatic coupling to a large Josephson junction, Phys. Rev. B \textbf{70}, 224515 (2004).
\bibitem{kk5} F. Y. Zhang, C. P, Yang, X. L. He, H. S. Song, Generation of macroscopic entangled coherent states with large Josephson junctions, Phys. Lett. A \textbf{378}, 1536 (2014).
\bibitem{kk6} X. L. He, Y. Liu, J. Q. You, and F. Nori, Variable-frequency-controlled coupling in charge qubit circuits: Effects of microwave field on qubit-state readout, Phys. Rev. A \textbf{76}, 022317 (2007).
\bibitem{kk6} Y. Kubo, F. R. Ong, P. Bertet, D. Vion, V. Jacques, D. Zheng,
A. Dr\'{e}au, J. F. Roch, A. Auffeves, F. Jelezko, J. Wrachtrup, M. F.
Barthe, P. Bergonzo, and D. Esteve, Strong Coupling of a Spin Ensemble to a Superconducting Resonator, Phys. Rev. Lett. \textbf{105}, 140502
(2010).
\bibitem{m} S. B. Zheng and G. C. Guo, Efficient Scheme for Two-Atom Entanglement and Quantum Information Processing in Cavity QED, Phys. Rev. Lett. \textbf{85}, 2392 (2000).
\bibitem{w1} E. Sj\"{o}qvist, D. M. Tong, L. M. Andersson, B. Hessmo, M.
Johansson, and K. Singh, Non-adiabatic holonomic quantum
computation, New J. Phys. \textbf{14}, 103035 (2012).
\bibitem{w2} A. A. Abdumalikov, , J. M. Fink, K. Juliusson, M. Pechal, S.
Berger, A. Wallraff, and S. Filipp, Experimental realization of
non-Abelian non-adiabatic geometric gates, Nature (London)
\textbf{496}, 482 (2013).
\bibitem{w3} G. Feng, G. Xu, and G. Long, Experimental Realization of
Nonadiabatic Holonomic Quantum Computation, Phys. Rev.
Lett. \textbf{110}, 190501 (2013).
\bibitem{w4} C. Zu,W.-B.Wang, L. He,W.-G. Zhang, C.-Y. Dai, F.Wang, and
L.-M. Duan, Experimental realization of universal geometric
quantum gates with solid-state spins, Nature (London) \textbf{514}, 72
(2014).
\bibitem{w5} S. Arroyo-Camejo, A. Lazariev, S. W. Hell, and G. Balasubramanian,
Room temperature high-fidelity holonomic singlequbit
gate on a solid-state spin, Nat. Commun. \textbf{5}, 4870
(2014).
\bibitem{w6}
Shi-Biao Zheng, Chui-Ping Yang, and Franco Nori, Comparison of the sensitivity to systematic errors between nonadiabatic non-Abelian geometric
gates and their dynamical counterparts, Phys. Rev. A \textbf{93}, 032313 (2016).
\bibitem{w7} Zheng-Yuan Xue, Jian Zhou, and Yong Hu, Holonomic Quantum Computation with All-Resonant Control in Circuit Quantum Electrodynamics, arXiv:1601.07219v2.
\bibitem{n} S. Haroche, in \emph{Fundamental Systems in Qunantum Optics}, edited
by J. Dalibard and J. Zinn-Justin (Elsevier, New York, 1992), p.
767.
\bibitem{d1} A. Blais, J. Gambetta, A. Wallraff, D. I. Schuster, S. M. Girvin, M. H. Devoret, and R. J. Schoelkopf, Quantum-information processing with circuit quantum electrodynamics, Phys. Rev. A \textbf{75}, 032329 (2007).
\bibitem{p} S. B. Zheng, Quantum-information processing and multiatom-entanglement engineering with a thermal cavity, Phys. Rev. A \textbf{66}, 060303 (2002).

\bibitem{o1} C. P. Yang, Y. X. Liu, and F. Nori, Phase gate of one qubit simultaneously controlling n qubits in a cavity, Phys. Rev. A \textbf{81}, 062323 (2010).
\bibitem{m1} J. Harrison, M. J. Sellars, and N. B. Manson, Measurement of the optically induced spin polarisation of N-V centres in diamond, Diam. Relat. Mater. \textbf{15}, 586 (2006).
\bibitem{m2} G. Balasubramanian, P. Neumann, D. Twitchen, M. Markham, R. Kolesov, N. Mizuochi, J. Isoya, J. Achard, J. Beck, J. Tissler, V. Jacques, P. R. Hemmer, F. Jelezko, and J. Wrachtrup, Ultralong spin coherence time in isotopically engineered diamond, Nat. Mater. \textbf{8}, 383 (2009).
\bibitem{m22} N. Bar-Gill, L. M. Pham, A. Jarmola, D. Budker, R. L. Walsworth, Solid-state electronic spin coherence time approaching one second, Nature Commun. \textbf{4}, 1743 (2013).
\bibitem{w} J. F. Poyatos, J. I. Cirac, and P. Zoller, Complete Characterization of a Quantum Process: The Two-Bit Quantum Gate, Phys. Rev. Lett. \textbf{78}, 390 (1997).
\bibitem{ww} M. J. Tao, M. Hua, Q. Ai, and F. G. Deng, Quantum-information processing on nitrogen-vacancy ensembles with the local resonance assisted by circuit QED, Phys. Rev. A \textbf{91}, 062325 (2015).
\end{references}
\end{document}